\documentclass[tighten]{aastex6}

\usepackage{epsfig,graphicx,amssymb,amsmath,epstopdf,mathrsfs}
\usepackage{subfigure}

\def\arcdeg{\hbox{$^\circ$}}

\def\simlt{\mathrel{\hbox{\rlap{\hbox{\lower4pt\hbox{$\sim$}}}\hbox{$<$}}}}
\def\simgt{\mathrel{\hbox{\rlap{\hbox{\lower4pt\hbox{$\sim$}}}\hbox{$>$}}}}

\renewcommand*{\thefootnote}{\fnsymbol{footnote}}

\begin{document}

\title{On the Fermi GBM event 0.4 sec after GW\,150914}

\author{J. Greiner\altaffilmark{1,2}, J.M. Burgess\altaffilmark{3,4}, 
       V. Savchenko\altaffilmark{5}, H.-F. Yu\altaffilmark{1,2}
} 

\email{jcg@mpe.mpg.de, jamesb@kth.se, savchenk@apc.in2p3.fr, sptfung@mpe.mpg.de 
      }

\altaffiltext{1}{Max Planck Institute for Extraterrestrial Physics,
    Giessenbachstrasse, 85748 Garching, Germany}

\altaffiltext{2}{Excellence Cluster Universe, Technische Universit\"{a}t 
      M\"{u}nchen,  Boltzmannstra{\ss}e 2, 85748, Garching, Germany}

\altaffiltext{3}{Oskar Klein Centre for Cosmoparticle Physics, SE-106\,91 
    Stockholm, Sweden}

\altaffiltext{4}{Dept. of Physics, KTH Royal Institute of Technology, AlbaNova,
     SE-106\,91 Stockholm, Sweden}

\altaffiltext{5}{Francois Arago Centre, APC, Universit\'{e} Paris Diderot,
           CNRS/IN2P3, CEA/Irfu, Observatoire Paris, Sorbonne Paris Cit\'{e},
           10 rue Alice Domon et L\'{e}onie Duquet, 75205 Paris Cedex 13,
           France}

\shorttitle{On the GBM-GW\,150904 event }


\begin{abstract}
In view of the recent report by \citet{2016arXiv160203920C}
we analyse continuous TTE data of Fermi-GBM around the time of the 
gravitational wave event GW\,150914. We find that after proper accounting
for low count statistics, the GBM transient event at 0.4 s after 
GW\,150914 is likely not due to an astrophysical source, but consistent 
with a background fluctuation, removing the tension between the INTEGRAL/ACS 
non-detection and GBM. Additionally, reanalysis of other short GRBs shows
that without proper statistical modeling the fluence of faint events is 
over-predicted, as verified for some joint GBM-ACS detections of 
short GRBs. We detail the statistical procedure to correct these biases.
As a result, faint short GRBs, verified by ACS detections, 
with significances in the broad-band
light curve even smaller than that of the GBM-GW150914 event are recovered as
proper non-zero source, while the GBM-GW150914 event is consistent with
zero fluence.
\end{abstract}


\maketitle


\section{Introduction}
\label{sec:intro}

The gravitational wave (GW) era started with an overwhelmingly strong  and 
convincing event (called GW\,150914), reported to originate from an unexpected
source of a massive black hole binary merger \citep{2016PhRvL.116f1102A}. 
Neither were merger rates of such massive binaries 
well established \citep[but see][for optimistic rates]{Belczynski:2010}, 
nor is it clear whether or not one should expect an electromagnetic signal 
\citep{Perna:2016, Lyutikov:2016}.
Finding the electromagnetic counterpart is the next challenge
which is difficult also because of the large error circles
of the present GW detection capability \citep{Abbott:2016}.

With the all-sky gamma-ray burst monitor (GBM) onboard the Fermi
satellite a flaring event with a signal-to-noise ratio of 5.1 and a 
false alarm probability of 0.2--2.8\% (depending on the assumed 
likelihodd distribution over the search window)
 was reported \citep{2016arXiv160203920C}.
If true, and indeed related to the GW event,  this would be a
fascinating discovery, allowing truly astrophysical constraints on the
massive black hole binary merger.  In view of this importance of
correlated electromagnetic signals to gravitational wave events,
particular scrutiny is required in the evaluation of candidate
electromagnetic signals.  Here, we report some additional analysis of
the Fermi/GBM data.

\begin{figure}[th]
  \centering
  \includegraphics[width=11.2cm]{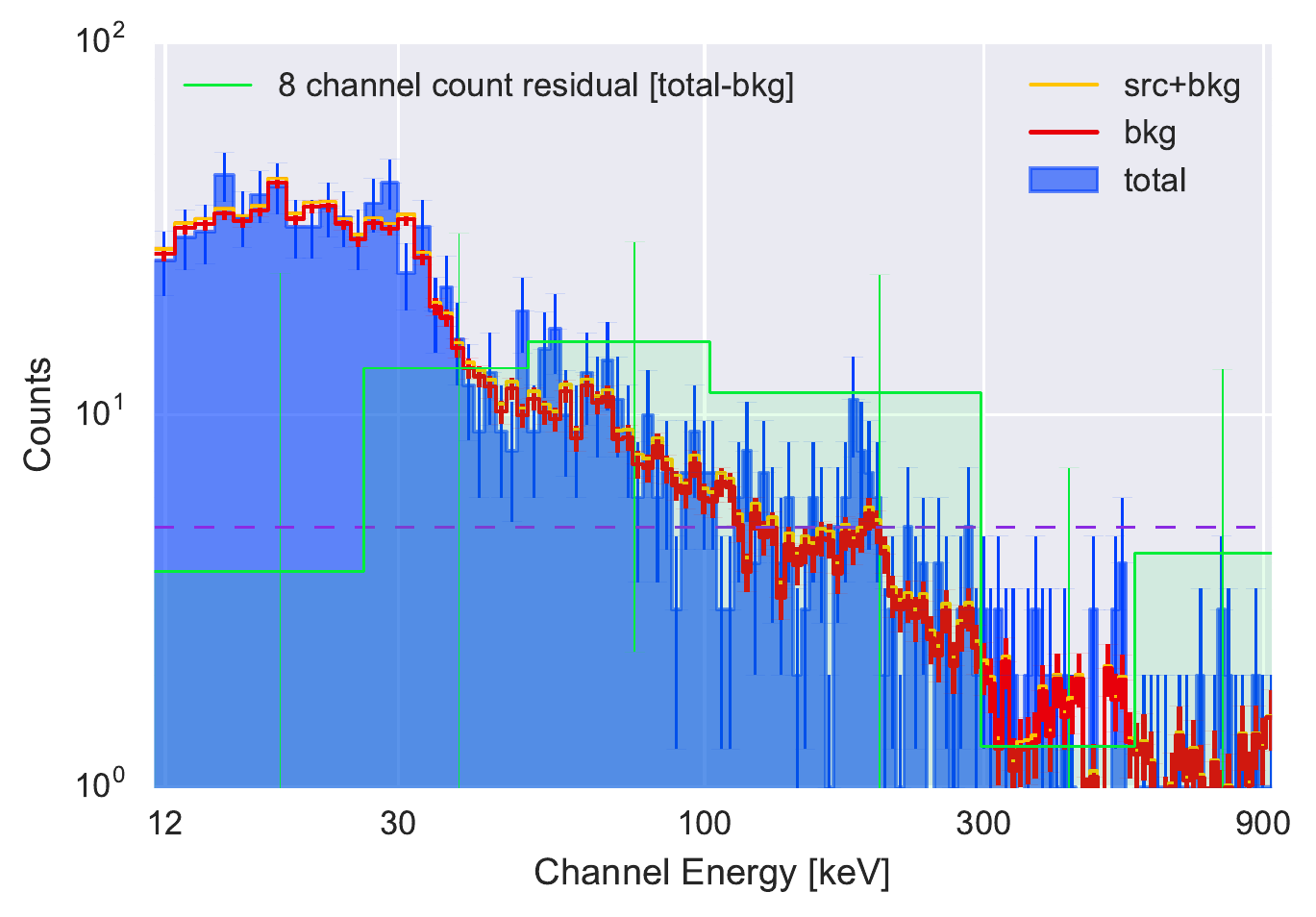}
  \caption{Spectral distribution of counts for the GBM event between
    0.384-1.402 s in NaI 5.  Shown are the total raw counts
    (blue), the background model from our fitted polynomial (red), the
    background plus source model (yellow) using the spectral
    parameters from our fit, and the residual source counts (green)
    rebinned into exactly the 8-channel spectrum as used in
    \citet{2016arXiv160203920C}, with the lowest (4-8 keV) and highest
    (overflow, i.e. $>$1 MeV) channel omitted.
     The highest-count
    channel is the one at 50--100 keV with 13 counts, demonstrating
    the low-count regime of the spectrum. The purple dashed
      line indicates the level at which the $\chi^2$ background
      fitting method breaks down.
    The blue and green error bars show the \cite{Skellam1946} 
    confidence intervals.
  \label{fig:countspec}}
\end{figure}

We approach our analysis in two steps. First, we examine
  methodology implemented in the standard GBM analysis tools and
  improve upon it paying special attention to the statistical nature
  of the data in light of the fact that the alleged source is weak and
  background dominated (Fig. \ref{fig:countspec}, Sect. 2). 
  Next, we repeat the spectral analysis
  following the approach of \citet{2016arXiv160203920C} using the
  standard tools for GBM analysis
  as well as examine the surrounding
  temporal background spectrum (Sect. 3). Finally, we implement a more
  statistically advanced approach to the spectral analysis which leads
  us to the conclusion that the source spectrum is more consistent
  with the background, rather than a short-hard GRB (Sect. 4).
  After demonstrating consistency with the INTEGRAL/ACS non-detection (Sect. 5),
  we conclude that caution should be taken with interpreting
  the GBM event as an astrophysical source.

\section{GBM data analysis}
\label{sec:GBM}
  The GBM is a non-imaging instrument relying on temporal
  isolation of signal against background. In practice, an observer
  identifies a source time interval in the light curve by its
  transient nature, compared to the smoothly-varying behavior of the
  background. Sophisticated algorithms may be applied to identify this
  source time interval, i.e., Bayesian blocks \citep{Scargle:2013},
  signal-to-noise threshold, etc. Regardless, once a source time
  interval is identified, the background must be modeled so that it
  can be properly accounted for in spectral fitting of the source
  interval. Therefore, GBM spectral analysis consists of a two-step
  process, temporal fitting of the background lightcurve and
  subsequent fitting of the source spectra. We will detail both
  processes in the following two sub-sections.

As a starting note, we recall that the event reported by
\cite{2016arXiv160203920C} was not identified by any of the standard
undirected offline
GBM search procedures, due to its faintness. 
Instead, it was identified by a targeted search, seeded with the
time of the GW event \citep{2015ApJS..217....8B}.
Indeed, the count distribution
of the event is fully consistent with the noise distribution before
and after the event. Two numbers given in \cite{2016arXiv160203920C}
are particularly worth noting: \\
(i) One is the signal-to-noise ratio SNR = 5.1 (corresponding to
 a significance of $\simlt\,3\sigma$) as given in their Fig. 2.
 This estimate is not determined purely on raw counts, but uses
 a spectral model (chosen to be a Band function with a low-energy
 power law of spectral slope 0 for the
 energy range where most counts are detected) to create, for a given
 sky location, a model rate per energy channel which then is used 
 to weight the raw counts of the different detectors for the summing
 towards a 
 ``signal-to-noise optimized light curve'' \citep{2016arXiv160203920C}.
While this hard spectral model best matched the observed counts
 in the GBM detectors \citep{2016arXiv160203920C}, we show below
 that a more appropriate spectrum has a spectral index of $-1.8$, thus
 lowering the SNR. \\
(ii) The other important number is the false alarm probability of 0.2\%
 \citep{2016arXiv160203920C}  which corresponds to 2.9$\sigma$, 
which is calculated from the measured false alarm rate of spectrally 
hard events in 2.5 days of GBM data, 
while that of spectrally  soft events 
is larger (see Fig. 3 in \cite{2016arXiv160203920C}).  
Due to this reason, the given false alarm probability of 
 0.2\% is an optimistic lower limit.

\begin{figure}[th]
  \begin{center}
  \includegraphics[width=10.2cm]{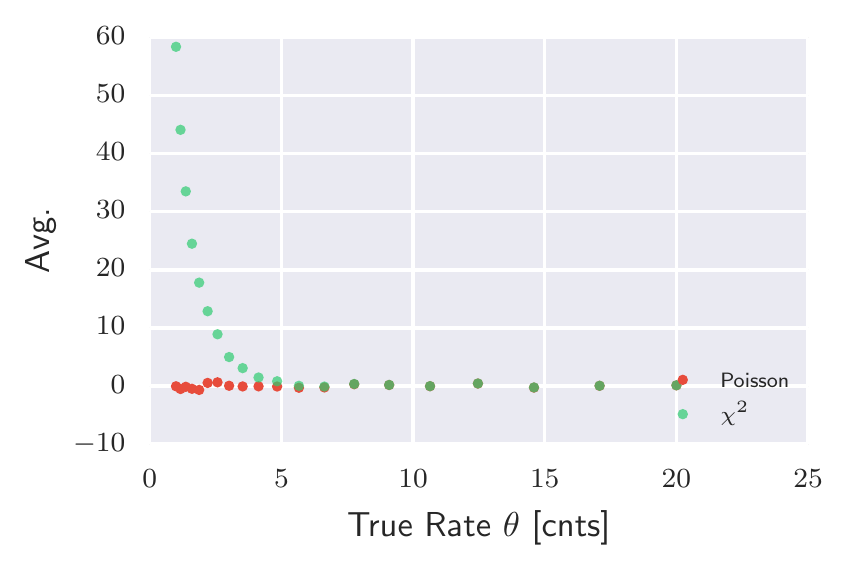}
  \caption{The average percent difference between the simulated and 
   recovered $\theta$ as a function of the  true rate
   for both the Poisson and $\chi^2$ background fitting methods.
   While the modification of $\chi^2$ made in RMFIT 
   (weighted $\chi^2$) improves the 
   results, it still a poor statistic at very low-counts which frequently 
   occur in high-energy channels. \\
  \label{fig:bkg2}}
  \end{center}
\end{figure}

\subsection{Background Behaviour}
\label{sub:bkg}

Due to the nature of GBM backgrounds, it is difficult to define a
spectral background model. The background contains a superposition of
astrophysical, terrestrial, solar, and instrumental sources, 
each with its own time variability.  Therefore, an assumption is made
that in each GBM energy channel, the background is smoothly varying
with time. With this assumption one selects a time interval before and
after the source time interval and fits a polynomial in time to the
count evolution to these time intervals but excluding the source time
interval. This polynomial is then interpolated through the source time
interval to estimate the background counts in each energy channel
during the time of source activity.  A major assumption of this
approach is that the background does not vary during the source time
interval more than the smooth nature apparent in the surrounding
background time interval. This assumption is most likely valid if the
source time interval is short compared to the time scale of the
variations of the fitting polynomial.

To determine the polynomial shape of the temporal background count evolution,
a fit to statistically fluctuating data must be performed. Under the assumption
that these counts should
be Poisson distributed, the statistic to choose for maximum-likelihood
estimation (MLE) is the Poisson likelihood \citep[e.g. the so-called Cash 
statistic,][]{Cash:1979}:

\begin{equation}
  \label{eq:1}
   L = \prod_{i=1}^{N}\frac{  M_i^{S_i} e^{-M_i}}{S_i!}
\end{equation}

\noindent where $M_i$ and $S_i$ are the model and detected counts in the
$i^{th}$ of $N$ bins, respectively. Taking the $-2 \log L$ and
approximating the factorial term with Stirling's approximation 
\citep{Abram:2002}\footnote{We note that this final form is sometimes
  referred to as the Castor statistic. However, it is simply the full
  Poisson likelihood, and its likelihood ratio asymptotically
  approaches a $\chi^2$ distribution under the condition that the
  Fisher information matrix is positive definite, i.e., the model is linear 
  in its parameters \citep{Wilks:1938}. Often, the
  terms dependent only on the data are dropped as they do not affect
  minimization \citep{Cash:1979}.}, we have:
\begin{eqnarray}
  \label{eq:2}
   -2\log L & \approx & 2\sum_{i=1}^{N} M_i - S_i \log M_i + S_i \log S_i - S_i \\
        & = & 2 \sum_{i=1}^{N} M_i -S_i + S_i (\log S_i - \log M_i) \text{.}
\end{eqnarray}

\noindent The polynomial coefficients can be estimated via minimizing the
log-likelihood of the background data and the polynomial integrated
over each time bin.

In the RMFIT\footnote{\url{http://fermi.gsfc.nasa.gov/ssc/data/analysis/rmfit/}}
software, generally used for GBM GRB data, the approach
for determining the background is similar to the above method, except
a weighted least-squares ($\chi^2$) statistic is chosen to fit the
polynomial coefficients via a two-pass method. First, the polynomial
coefficients are estimated by solving the least-squares linear
equations assuming that the data variance is equal to the number of
counts. Then, the fit is performed again, but using the variance of
the model determined in the first pass. This is intended to compensate
for the fact that at low counts, the Gaussian approximation of the
Poisson distribution is poor.

In the high-count regime, this approach may be valid.
However, due to the fact that a polynomial is fit in each
  energy channel and that the background has fewer counts at high
  energies, the $\chi^2$ statistic can be very biased
  \citep{BakerCousins84, Humphrey:2009, Andrae:2010}.  
  To estimate this effect, we
  simulate synthetic light curves obeying a zero-order polynomial
  evolution in time over a period of 20 seconds. We choose a set of
  levels for the background mean rate ($\theta$) ranging from 1-100
  cts s$^{-1}$. For each $\theta$, we simulate 500 lightcurves, and
  fit a polynomial to them with two different statistics criteria:
  Cash, and $\chi^2$. We find that the Cash statistic well recovers
  the true $\theta$ even at very low counts, while both $\chi^2$
  methods have a large bias at low counts (see Fig.  \ref{fig:bkg2}).
Therefore, we will use the Poisson likelihood to fit the background in
our spectral analysis, but will compare the results to what is found
using RMFIT and $\chi^2$.

\subsection{Spectral Fitting Statistics}

In the low source count regime, the choice of spectral
  fitting statistic is crucial. The total count rate in a source
  interval is distributed as a Poisson process, but we have more
  information via the background fit to better constrain the spectral
  fit parameters. The total counts in a spectral bin ($S_i$), is
  composed of source and background ($B_i$). Since we have estimated
  the background via a polynomial model, we can include this
  information in the Poisson likelihood. Let $P(t;\vec{\theta})$ be
  the estimated polynomial background rate model in a given energy
  channel where $t$ is time and $\vec{\theta}$ are the polynomial
  coefficients. Then, for a given source interval from time $T_1$ to
  $T_2$, we have
\begin{equation}
  \label{eq:3}
  B_i = \int_{T_1}^{T_2}P(t;\vec{\theta}) {\rm d}t \text{.}
\end{equation}

\noindent Now, $\vec{\theta}$ are fitted coefficients with associated
statistical errors and therefore $B_i$ also has statistical error
($\sigma_B$). Using the covariance matrix for $\vec{\theta}$, we can
estimate $\sigma_B$ via standard Gaussian error
propagation\footnote{Assuming that the error distribution of the
  coefficients is Gaussian which should be safe. If they are not then
  we cannot employ standard error propagation without Monte Carlo
  methods.} \citep[see][for a review of standard Gaussian error
propagation]{Barlow:1993}. Error propagation assuming that the resulting error
distribution is Gaussian leads to Gaussian uncertainties on the
model estimated $B_i$ quite naturally\footnote{The fact that the
  estimated background model has a Gaussian error distribution should
  not be confused with the fact that the background data is Poisson
  distributed.}.

We must use a likelihood for Poisson data with a Gaussian background 
to fit the spectral data. This is embodied in the PG-statistic 
\citep{Arnaud:2015}:
\begin{equation}
  \label{eq:5}
  -2 \log L = 2 \sum_{i=1}^N M_i + t_s f_i -S_i \log(M_i t_s f_i)+ \frac{1}{2 \sigma_{B,i}^2} (B_i- t_s f_i)^2 - S_i(1-\log S_i) \text{.}
\end{equation}
Here, $t_s$ is the source interval duration and $f_i$ is the
profiled-out background model\footnote{see \url{https://heasarc.gsfc.nasa.gov/xanadu/xspec/manual/XSappendixStatistics.html} for details.}. This is the
statistic we will minimize in our spectral analysis.

In the RMFIT software, the background is assumed to be perfect with no
variance. Therefore, only the total counts are assumed to have
statistical variability. To deal with this, RMFIT modifies the Poisson
likelihood by adding the estimated background from Equation \ref{eq:3}
to the estimated spectral model counts:

\begin{equation}
    \label{eq:cstatmod}
    -2 \log L = 2\sum_{i=1}^{N}(M_i+B_i))-S_i +S_i(ln(S_i) - ln(M_i+B_i)) ) \text{.}
\end{equation} 

\subsection{Method comparison: spectral simulations}
\label{sub:sim}

Since it is possible that the method of \citet{2015ApJS..217....8B} 
recovers a significant fluctuation above the typical background,
it is important to understand if the source has the
spectral parameters reported in \citet{2016arXiv160203920C}.
Therefore, we simulate the problem. We fit the background of
the data before the trigger over a duration of 10 s to obtain the
average background rate in each channel. We then simulate 10 s
background from this sampled spectrum assuming each channel has
Poisson fluctuations with their means derived from the sampled
spectrum. Next, we simulate a one second source with a power law
spectrum. We choose the power law index to be --1.4 and three different
amplitudes: A=${2\cdot10^{-3},\cdot10^{-4},0.0}$. 
The first amplitude
models the claim of \citet{2016arXiv160203920C},  the second simulates 
an order of magnitude weaker source, and finally the last amplitude 
implies no source.
 For each set of
spectral parameters, we generate 100 sets of source and background
simulations and create TTE data via the method described in
\citet{Burgess:2014fl}.

\begin{figure*}[ht]
\centering
\includegraphics[width=15.0cm]{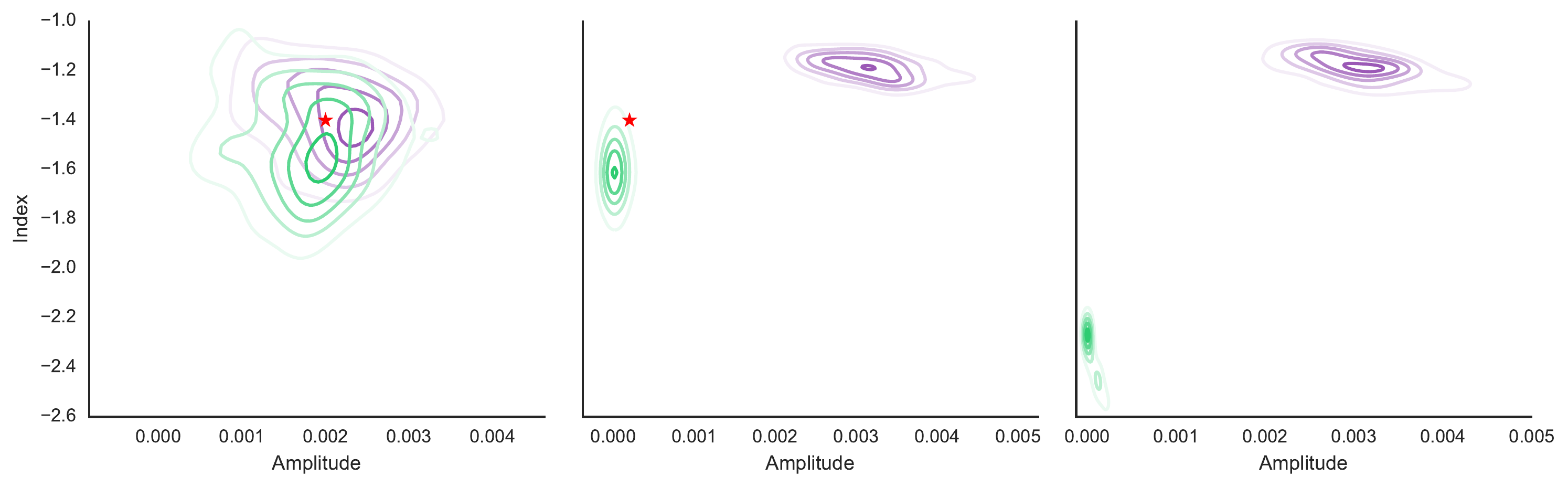}
\caption{Recovered parameters from simulated spectra of a power law
  with index --1.4 and assuming the source amplitude is 0.002 ph
  cm$^{-2}$ $s^{-1}$ (left panel), 0.0002 (middle) or zero (no source;
  right). The results of RMFIT are shown in purple and our MLEfit in
  green. The true parameters are indicated with a red star.}
  \label{fig:amp1}
\end{figure*}

For each simulation, we fit the simulated source spectra with three
different methods. First, we fit the background and spectrum in RMFIT. 
Next, we use the Poisson method to fit the background,
and fit the source spectra with the  standard PGStat statistic.
To minimize the
likelihood, we use MINUIT and our own RSP matrix convolution software
for computational simplicity (hereafter MLEfit), but verified our
results by producing PHA files for source and background and fitting
them with XSPEC\footnote{Unfortunately, XSPEC does not have the
  modified C-stat used in RMFIT or we could perform our full analysis
  with XSPEC.}. We effectively compare three different methods for
fitting spectra. 
Figure \ref{fig:amp1} shows the results of the simulations
where we assume the same source spectrum reported in
\citet{2016arXiv160203920C}.

MLEfit using PGStat recovers the true simulated amplitude more
accurately than any other method. 
The same has been shown to be true for a simulation of different
spectral slopes at identical amplitudes \citep[see Fig. 7.3. in][]{Arnaud:2011}.
RMFIT is biased to slightly higher amplitude
values. When the source amplitude is reduced by an order of magnitude,
MLEfit is about to find the correct amplitude with either statistic
though the modified C-stat is slightly biased towards higher values. 
On the other hand, RMFIT recovers a source amplitude
resembling what is found in \citet{2016arXiv160203920C} (see Figure
\ref{fig:amp1}).
 Similarly, when there is no source, MLEfit finds a source
amplitude near zero, while RMFIT still finds a source amplitude of
$\sim 3\cdot 10^{-3}$. This has two implications. MLEfit is able to
distinguish between source and background while RMFIT is including
background in its source spectrum. 
The fact that RMFIT finds a source
spectrum similar to what is reported in \citet{2016arXiv160203920C}
even when there is \textit{no source} indicates that the background
spectrum can be modeled with a power law having an amplitude and
overall spectrum very similar to the reported source spectrum in
\citet{2016arXiv160203920C}.
We further demonstrate this in the following section.

Combining the low count bias of $\chi^2$ which can underestimate the
background level and the assumption of no variance on this
underestimated background via the use of RMFIT's modified C-stat, it
is possible to find a source spectrum and flux that is unrealistically
too large as we will demonstrate using our approach to spectral
fitting.

\section{Analysis of the GBM event}

\subsection{Results with RMFIT}

Following \citet{2016arXiv160203920C}, we generate 10 sets of detector
response matrices (RSPs) corresponding to the 10 locations listed in
the same work, using the standard stand-alone response generator.
We use the GBM team created software, RMFIT to
bin the data, make energy selections equivalent with those in
\cite{2016arXiv160203920C}, as well as fit the background to the
surrounding count light curve. 
We fit a power-law spectrum to the data, using the 10
different RSPs discussed above. Table \ref{tab:fit} details these fits. 
Within the statistical errors, these are compatible with the
spectra derived by \cite{2016arXiv160203920C}, in particular the
details given in their sect. 3.2 and Fig. 5.

\begin{table}[hb]
\caption{GBM spectral parameters (and their 1$\sigma$ errors) 
 at 10 positions along the LIGO arc, derived with RMFIT and with PGStat.}
\centering
\begin{tabular}{cccccccc}
    \hline
    \noalign{\smallskip}
    \hline
    \noalign{\smallskip}
    RA  & DEC   & \multicolumn{3}{c}{RMFIT-based analysis} & \multicolumn{3}{c}{PGStat-based analysis} \\
        &       & Amplitude     & Index & Fluence & Amplitude~~ & Index & Fluence \\
  (deg) & (deg) & ~~~(ph/cm$^2$/s)~~~ &       & (10$^{-7}$ erg/cm$^2$)~~~  & (ph/cm$^2$/s) &       & (10$^{-7}$ erg/cm$^2$)\\
        &       & (@100 keV)    &       & (10--1000 keV)~~~ & (@100 keV)    &       & (10--1000 keV) \\
    \noalign{\smallskip}
    \hline
    \noalign{\smallskip}
 84.0 & -72.8 & 0.0043$\pm$0.0020 & $-1.44\pm0.14$ & 4.3$\pm$1.5 & ~~0.0035$\pm$0.0031 & $-1.85\pm0.86$ & 2.7$\pm$2.6 \\
155.3 & -43.2 & 0.0019$\pm$0.0006 & $-1.26\pm0.11$ & 2.1$\pm$0.6 & ~~0.0008$\pm$0.0005 & $-1.50\pm0.25$ & 0.8$\pm$0.5\\
102.0 & -73.9 & 0.0039$\pm$0.0016 & $-1.42\pm0.13$ & 3.9$\pm$1.2 & ~~0.0025$\pm$0.0020 & $-1.93\pm0.43$ & 1.9$\pm$1.6\\
118.3 & -72.9 & 0.0034$\pm$0.0013 & $-1.39\pm0.12$ & 3.5$\pm$1.0 & ~~0.0018$\pm$0.0019 & $-1.79\pm0.42$ & 1.4$\pm$1.3\\
132.0 & -70.4 & 0.0030$\pm$0.0011 & $-1.36\pm0.12$ & 3.2$\pm$0.9 & ~~0.0014$\pm$0.0014 & $-1.72\pm0.39$ & 1.1$\pm$1.0\\
140.9 & -66.6 & 0.0026$\pm$0.0009 & $-1.33\pm0.11$ & 2.9$\pm$0.8 & ~~0.0014$\pm$0.0011 & $-1.69\pm0.32$ & 1.1$\pm$0.9\\
147.5 & -62.5 & 0.0024$\pm$0.0008 & $-1.31\pm0.11$ & 2.7$\pm$0.7 & ~~0.0009$\pm$0.0007 & $-1.57\pm0.32$ & 0.8$\pm$0.6\\
151.2 & -58.0 & 0.0022$\pm$0.0007 & $-1.29\pm0.11$ & 2.5$\pm$0.7 & ~~0.0009$\pm$0.0006 & $-1.54\pm0.28$ & 0.8$\pm$0.6\\
153.4 & -53.1 & 0.0020$\pm$0.0007 & $-1.28\pm0.11$ & 2.4$\pm$0.6 & ~~0.0009$\pm$0.0006 & $-1.50\pm0.26$ & 0.8$\pm$0.6 \\
153.9 & -48.2 & 0.0019$\pm$0.0006 & $-1.27\pm0.11$ & 2.2$\pm$0.6 & ~~0.0009$\pm$0.0005 & $-1.49\pm0.25$ & 0.8$\pm$0.5\\
    \noalign{\smallskip}
    \hline
\end{tabular}
\label{tab:fit}
\end{table}

We note that we have used the full 128-channel TTE data instead of
the 8-channel TTE data used in \cite{2016arXiv160203920C}, because a proper 
MLE statistic is accurate
regardless of the binning as pointed out in \citet{Cash:1979}.

\subsection{Spectral Variations of the Background}
\label{sub:spec}

The source spectrum reported in \citet{2016arXiv160203920C} is 
more similar to the background (cosmic X-ray background plus Earth albedo
spectrum; \citealt{2008ApJ...689..666A}) than of short GRBs. 
In Section \ref{sub:sim}, we
demonstrated that RMFIT will effectively find a source spectrum even
when no source is present and is therefore fitting strong, positive
fluctuations of the background. To further assess this claim, we fit 1.024
second intervals before and after the tentative detection. We
iteratively select the intervals as source and fit the surrounding
light curve as background. We use the Cash statistic to fit the
background and MLEfit with PGStat to fit the source since we have
demonstrated that the RMFIT methods are insufficient for low-count
analysis. As an additional check, we also fit the spectra with RSPs
generated from each of the 10 locations used in \citet{2016arXiv160203920C}.

\begin{figure}[bh]
\includegraphics[width=8.2cm]{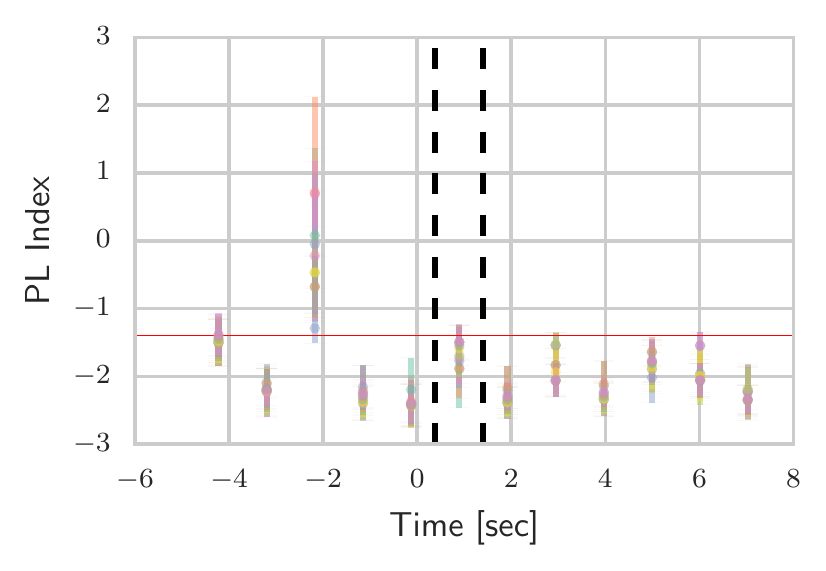}
\caption{Variation of spectral slopes of power-law 
spectra fitted to the residual flux after background subtraction.
The vertical dashed lines show the interval of the putative event,
and the red horizontal line the spectral index for the best-fit position
from  \citet{2016arXiv160203920C}.
 For each time interval, the power law
indices of all 10 positions along the LIGO localisation arc are shown.
\label{fig:bkg}}
\end{figure}

Figure \ref{fig:bkg} shows that the spectra of the surrounding time intervals
are very similar to the spectral parameters of the putative event,
regardless of position on
the sky used to generate the RSPs. Since the location is poorly
constrained, then the range of parameters covered by these plots could
be considered valid. Therefore, all recovered spectra are essentially
the same indicating that the source is most likely a strong positive
background fluctuation.

\section{Spectral Analysis with Proper Statistics}

\subsection{Spectral analysis with Cash and PGStat}

Here, we demonstrate the spectrum of the source candidate using
Cash statistics to fit the background, and PGStat to fit the
spectra. Table \ref{tab:fit} shows the best fit for each of the 
10 locations across the LIGO localisation arc. As can be seen, 
the amplitudes are much lower and the fluences reduced correspondingly. 
Also, the spectral slope is substantially softer.
It is still most likely that the source is a
strong background fluctuation, because the spectrum is 
not inconsistent with zero, and
consistent with what is expected from the combination of extragalactic 
background and Earth's albedo.
While the larger error in spectral slope still allows a hard spectrum,
the reality of these solutions is excluded by the INTEGRAL/ACS limits (see
end of this section, and the next one). 

\begin{figure}[htbp]
\begin{center}
\includegraphics[width=10cm]{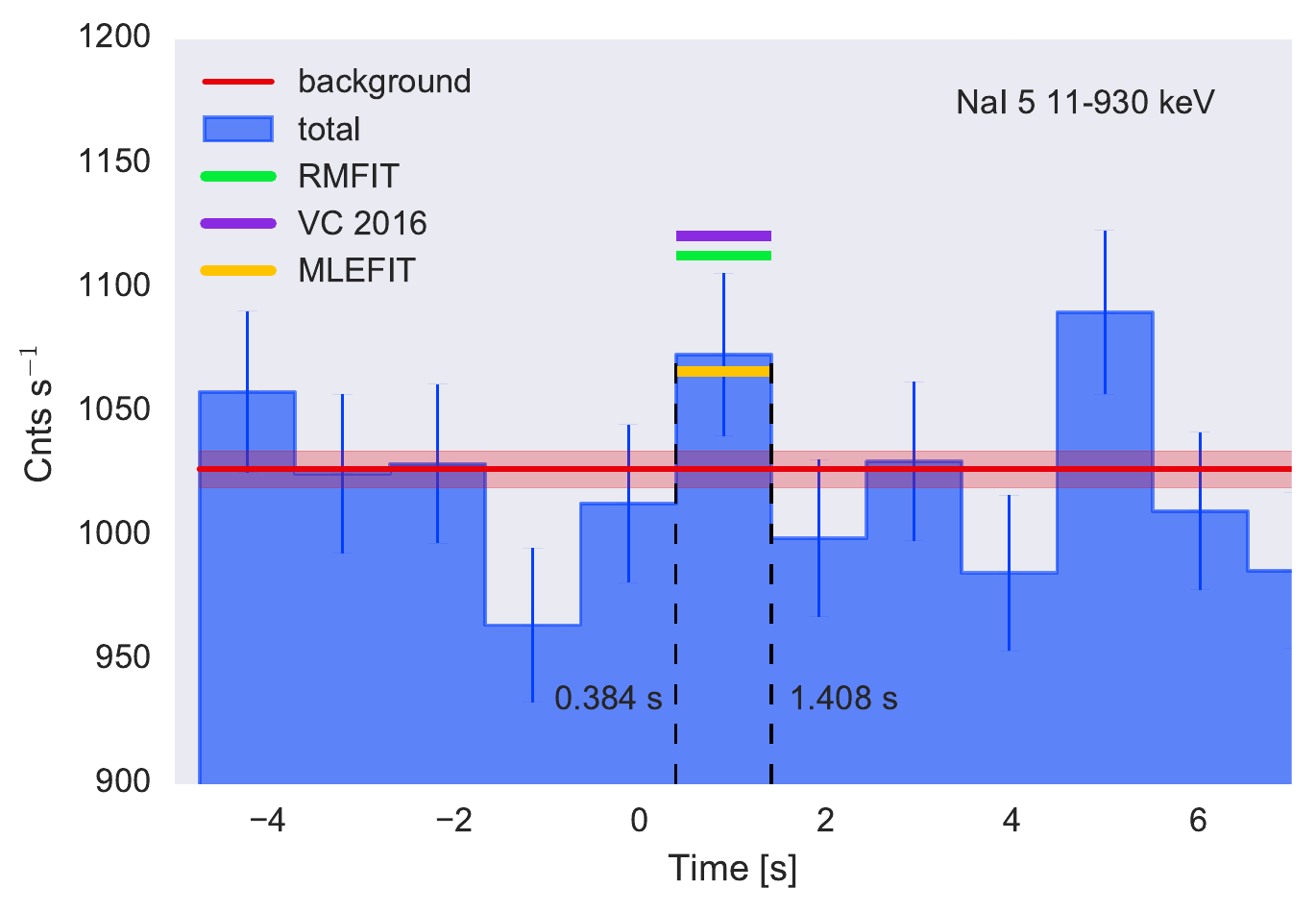}
\caption{The total, raw count light curve of NaI 5 (blue) integrated
  over 11--930 keV. The modeled background (red) with shaded 1$\sigma$
  Gaussian error is shown in red. 
Using the GBM DRM, we calculate the
  predicted counts from power law fits using our method (yellow), our
  fit with RMFIT (green) and the parameters reported in \citet{2016arXiv160203920C}.
Both methods that rely on RMFIT
  over-predict the expected counts. Additionally, it is easy to see
  that there are spikes in the raw light curve that are equally as
  bright as the alleged event.}
\label{fig:lc}
\end{center}
\end{figure}

Next, we examine the predicted counts from our spectral fits with
MLEfit, RMFIT, and those reported in
\citet{2016arXiv160203920C}. Figure \ref{fig:lc} shows the count light
curve of NaI 5 integrated over 11.5--930 keV. The background model
coming from the polynomial fit is shown with its 1$\sigma$ Gaussian
errors. The total Poisson error of the event is not too far
from the background, and indeed there are several peaks of similar
amplitude within this time slice. Using the RSP corresponding to 
RA = 155.3, Dec = -43.2 and the spectral parameters corresponding to the fits
at that location (see e.g. row 2 of Table \ref{tab:fit}),
we convolve the power-law photon model to produce
the predicted number of counts during the event. 
The counts predicted
from MLEfit are within the observed range of background subtracted
counts, but the results from our own RMFIT fits and those of
\citet{2016arXiv160203920C} (using RMFIT) over-predict the number of 
counts significantly.

\subsection{Bayesian approach}

As a secondary check, we attempt a Bayesian fit of the data
  where both the background polynomials in each channel and the
  spectra are fit simultaneously. This avoids the ambiguity of
  background error propagation and allows us to directly assess the
  background's effect on the spectral fit. The method will be detailed
  in \citet[][in prep.]{Burgess:2016}, however, we briefly detail
  the approach here. The full temporal and spectral model is defined
  as a piecewise function in time:

$\begin{displaystyle}
  \label{eq:lightcurve}
  f(t^i,\varepsilon;\theta_n^{j},\vec{\phi})= 
  \begin{cases}
    B(t^{i};\theta_n^{j}) \& t^j < t_{\rm start}\\
    B(t^{i};\theta_n^{j}) + S(t^{i},\varepsilon;\vec{\phi}) \& t_{\rm start} \leq t^j \leq t_{\rm stop}\\
    B(t^{i};\theta_n^{j}) \& t^j> t_{\rm stop}
  \end{cases}
\end{displaystyle}$

\noindent
where $t^i$ is the time in the $i^{\rm th}$ interval, $\theta_n^{j}$
is the $j^{\rm th}$-order polynomial coefficient in the $n^{\rm th}$
energy channel, $B$ is the background polynomial, and $S$ is the
source photon model with $\vec{\phi}$ parameters. The $ t_{\rm start}$ 
and $ t_{\rm stop}$ correspond to the source interval times. For the 
data at hand, we use a zero-order polynomial, $B$ in each channel and
again choose a power-law function for $S$. A log-uniform prior is
chosen for $\theta_n^{0}$ and the power law amplitude and a uniform
prior for the power law spectral index. Unlike the previous analysis, we are
fitting both the background and source simultaneously, so we can take both
the background and source as distributed via the Poisson process.

\begin{figure}[h]
  \centering
  \includegraphics[width=8cm]{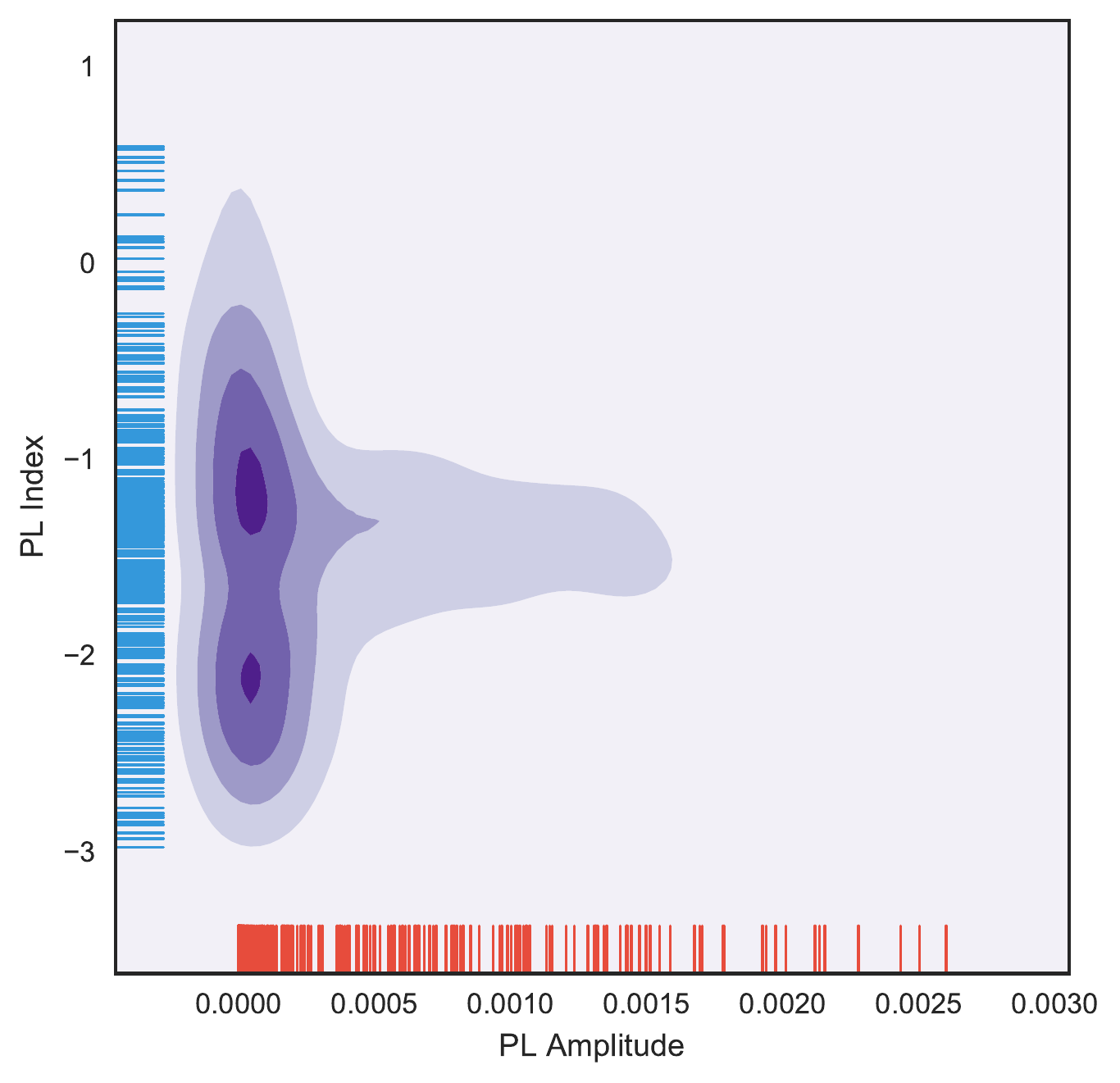}
  \caption{The posterior contours of the power law fit to the
    event. The blue and red rug plots indicate the distributions of
    the power law index and amplitude, respectively. The parameters are
    poorly constrained and not inconsistent with zero amplitude.}
  \label{fig:bayesreal}
\end{figure}

\begin{figure}[h]
  \centering
 \subfigure{\includegraphics[width=8.7cm]{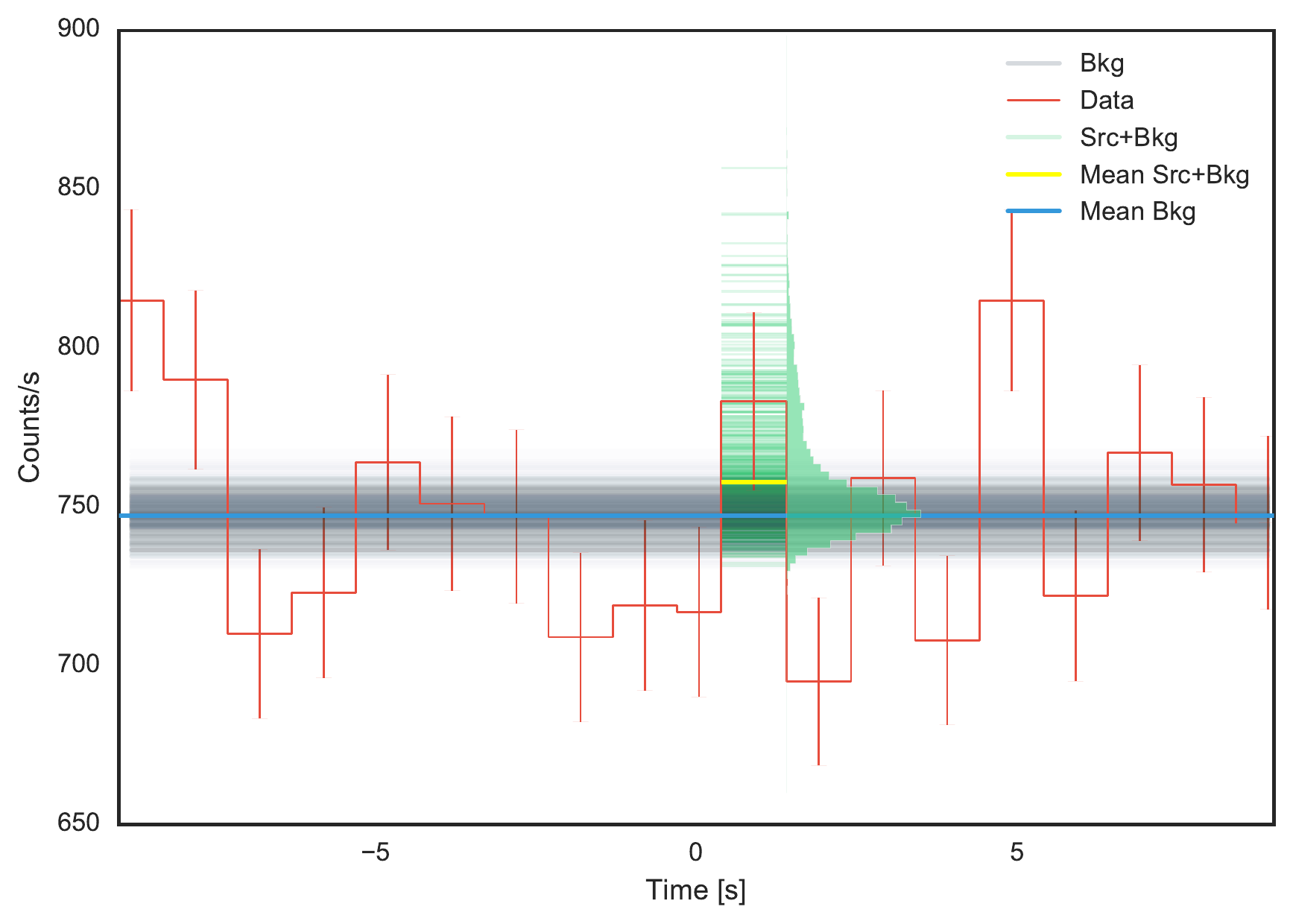}}\subfigure{\includegraphics[width=8.7cm]{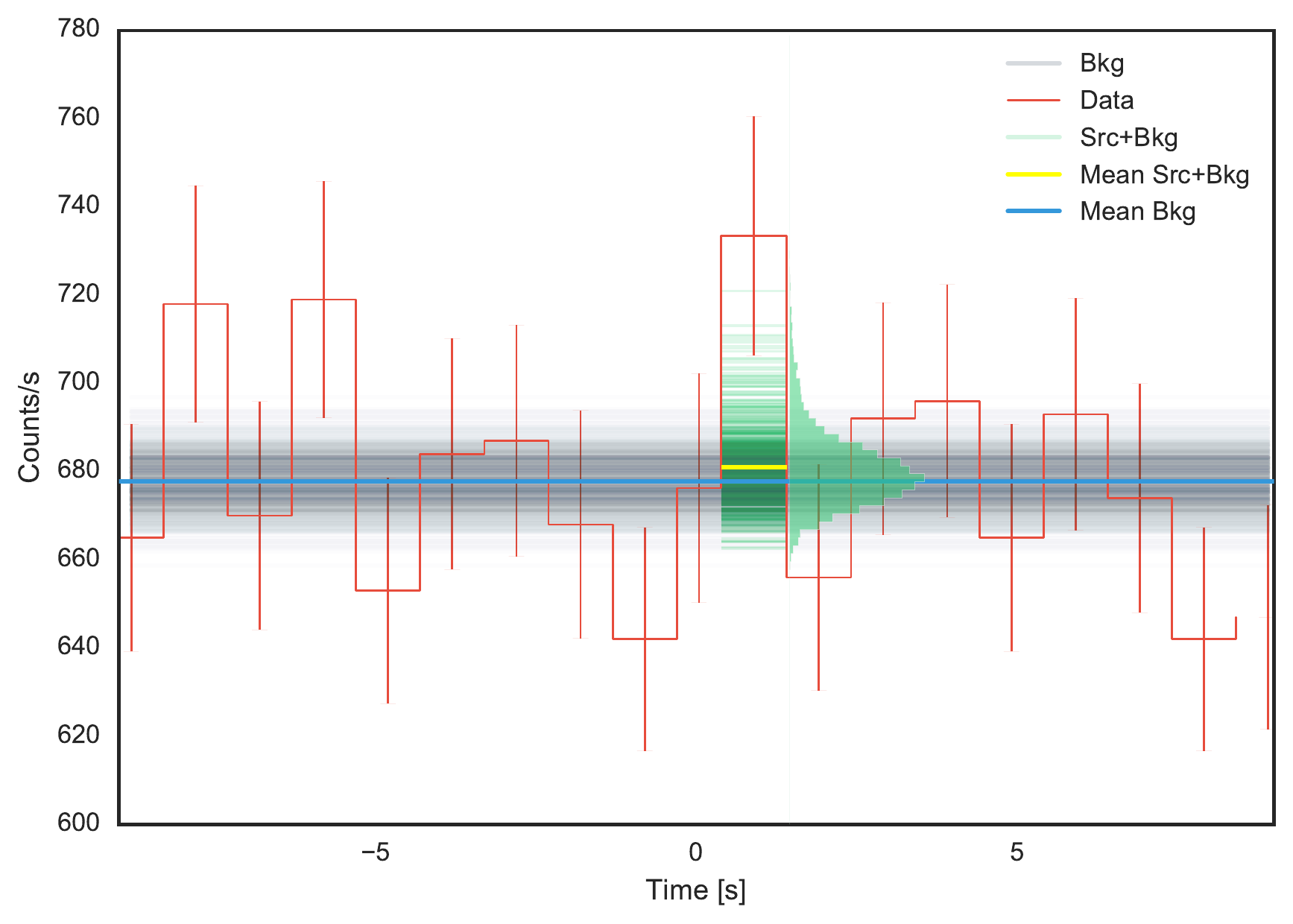}}
  \caption{The data from the NaI (left) and BGO (right) superimposed with the
background and source+background posteriors. The source+background posterior
(green histogram) clearly shows that the source is consistent with zero, i.e.,
background.}
  \label{fig:detreal}
\end{figure}

Figure \ref{fig:bayesreal} shows the resulting parameter posterior contours
indicating that the parameters are mostly unconstrained and the
amplitude is not inconsistent with zero. Moreover, Figure \ref{fig:detreal}
shows the light curves from the data with the background and
source + background posteriors for the NaI and BGO detectors. The
source + background histogram is clearly consistent with the
background only, suggesting that the source is a background
fluctuation and not an astrophysical source.

\begin{figure}[h]
  \centering
  \includegraphics[width=8cm]{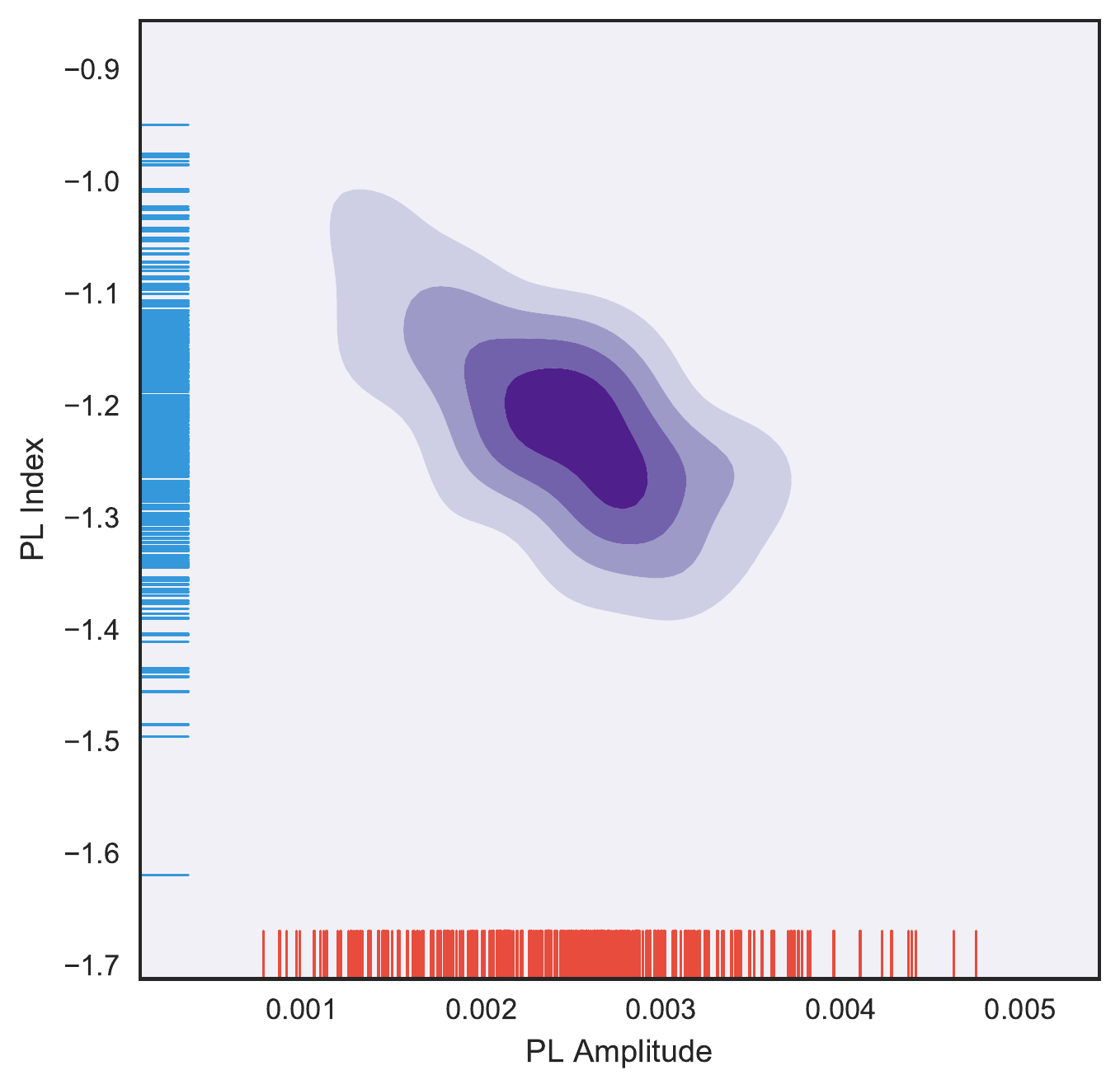}
  \caption{The same as Figure \ref{fig:bayesreal} but with simulated
    data with Amplitude=0.002 photons $s^{-1}$ cm$^{-2}$ and spectral
    index = $-1.4$. The parameters are well recovered at their simulated
    values. }
  \label{fig:bayessim}
\end{figure}

\begin{figure}[h]
  \centering
\subfigure{ \includegraphics[width=8.7cm]{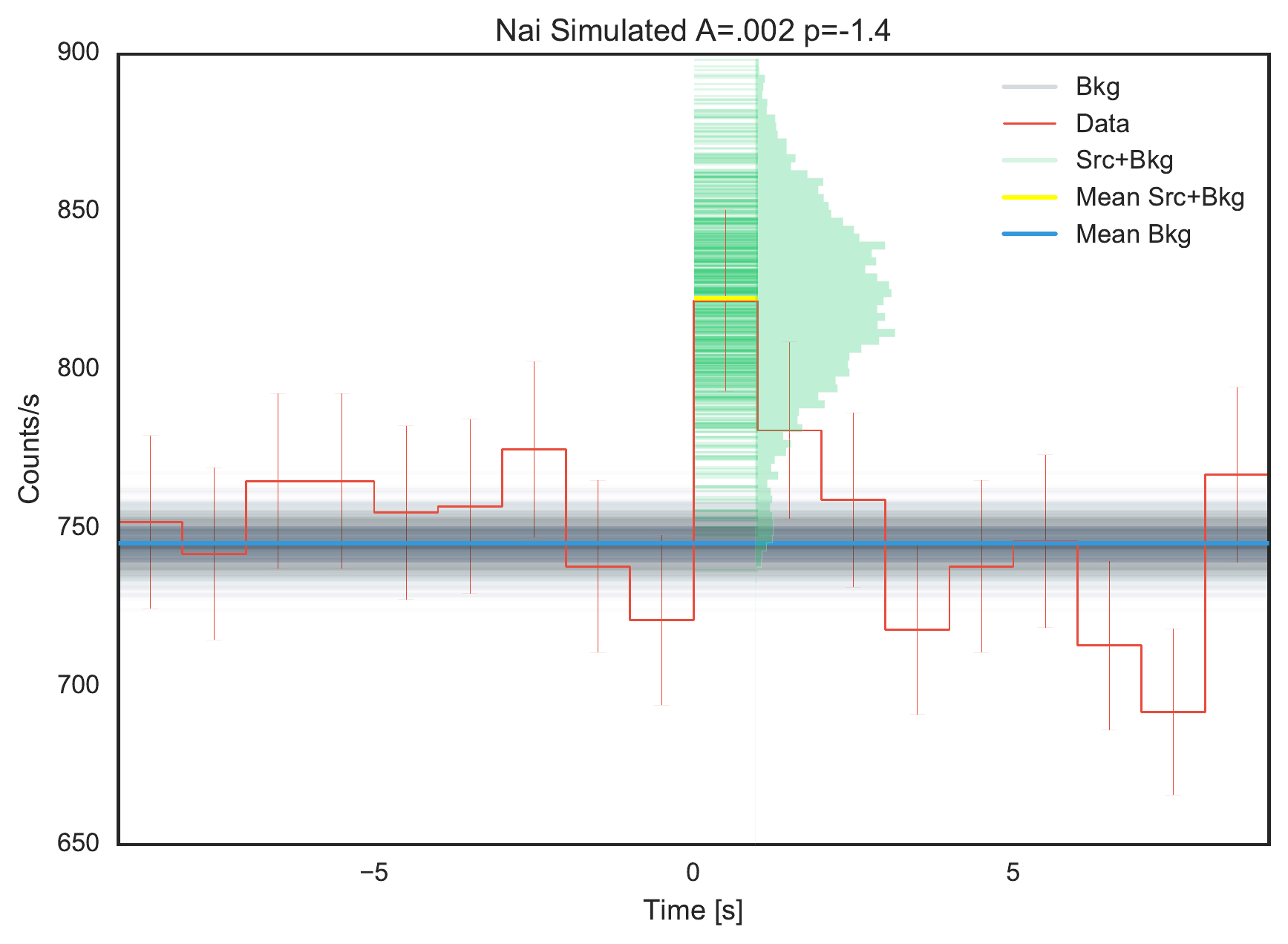}}\subfigure{\includegraphics[width=8.7cm]{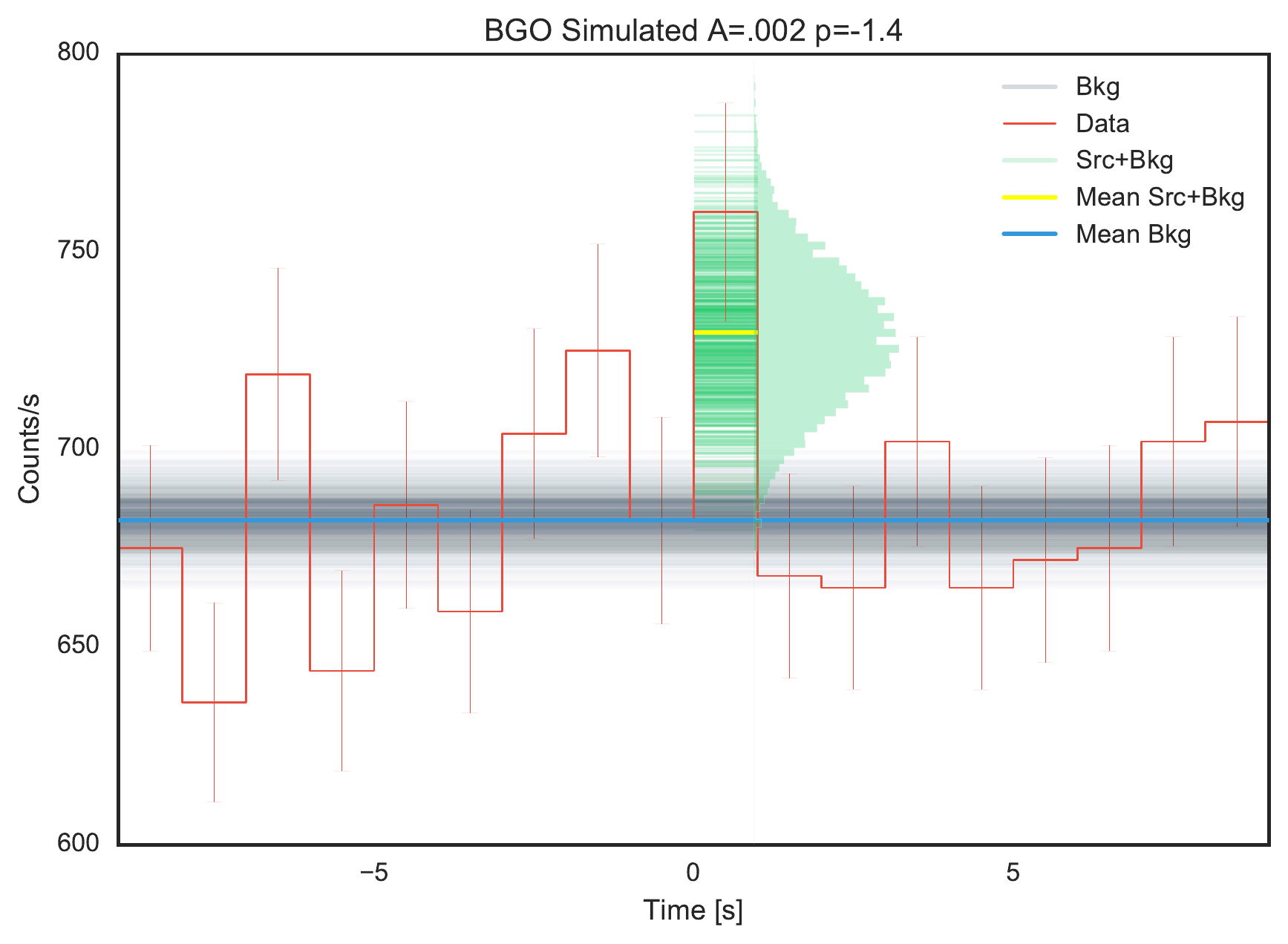}}
\caption{The same as figure \ref{fig:detreal} but for simulated
  data. The source+background posteriors (green histograms) clearly
  show that the source is not consistent with background as expected.}
  \label{fig:detsim}
\end{figure}

We again simulate the spectral parameters in
\cite{2016arXiv160203920C} to test if this method could recover the
parameters of a real source. Figures \ref{fig:bayessim} and \ref{fig:detsim}
show the results indicating that a true source with the reported 
spectral parameters would be recovered by the method.

To access the presence of an actual source, we use the Bayesian tool to fit the
data using the background polynomial only. Using the Monte Carlo samples from
the source + background fit and the background only fit, we compute the
deviance information criteria (DIC) \citep{Spiegelhalter:2002} for each model which
allows us to perform model comparison between them. The DIC compares model fit
and penalizes for  model complexity which can be a function of both data and
model. We find a DIC of 18341 and 18347 for background only and
source+background respectively. 
Choosing the lowest DIC implies that the
background only model is the best description of the data. 
This proposes an
interesting contrast to the claimed detection  in \cite{2016arXiv160203920C} via the
method of \cite{2015ApJS..217....8B}.
It is likely that the included information
in the simultaneous fit of the full resolution data is a better
characterization and this method correctly finds no significant source in
agreement with theory and the non-detection in all other observatories.

\subsection{Re-analysis of short GRBs}

Based on the previous analysis, we conjecture that the mismatch
in the resulting parameters depends on the statistical significance of the
residual source counts per event. In order to test this conjecture,
we re-analyzed with PGStat all 58 short GRBs from the GBM time-integrated 4-yr
catalog \citep{Gruber:2012}, which had the powerlaw model as BEST-fit model.
Fig. \ref{fig:deltas} shows the comparison of the resulting parameters
with those obtained using RMFIT (and listed in \citealt{Gruber:2012}).
We find that PGStat reproduces the fit parameters of RMFIT for strong GRBs,
while for faint GRBs the spectra are systematically softer and the fluences
smaller, exactly as shown in the above analysis of the GBM event.

The above analysis demonstrates the systematic effect of the differences
between the RMFIT package and the PGStat analysis. We can go one step further 
and ask, how the different best-fit parameters affect the prediction of
the detectability by INTEGRAL/ACS, and how these predictions compare
with the actual detections by INTEGRAL/ACS? Taking two short GRBs 
from the \citep{Gruber:2012} catalog which belong to the 20\% lowest fluence
group, we fold the spectral parameters (with their catalog sky position) 
through the INTEGRAL/ACS response and obtain predictions in the 
3--4$\sigma$ range.
Checking the INTEGRAL/ACS, we indeed find these GRBs at these confidence
levels, demonstrating that our PGStat analysis provides results which are
verified by detections with an independent instrument. These two short GRBs 
are added in Figures \ref{fig:deltas} and \ref{fig:ACSmap}. As previously,
RMFIT overpredicts the counts, and correspondingly the fluence in INTEGRAL/ACS.

Typical short GRBs have a spectral index of --1.4$\pm$0.2  \citep{Gruber:2012},
and our PGStat analysis of the faintest triggered short GRBs finds them
at indices as soft as $-1.6$ (with large errors, see Fig. \ref{fig:ACSmap}).
The best-fit spectral index of the GBM event close to GW150914 
GRB 120212353 is $-1.6$
(for a position outside the LIGO arc), and the spectral indices for positions
along the LIGO arc cover the range $-2.0$...$-1.5$ (also with large errors, 
see Fig. \ref{fig:ACSmap}). While there is some overlap between the two 
distributions, the GBM-GW150914 event is systematically softer than even the 
softest/faintest short GRBs.

\begin{figure}[htbp]
\begin{center}
\includegraphics[width=8.7cm]{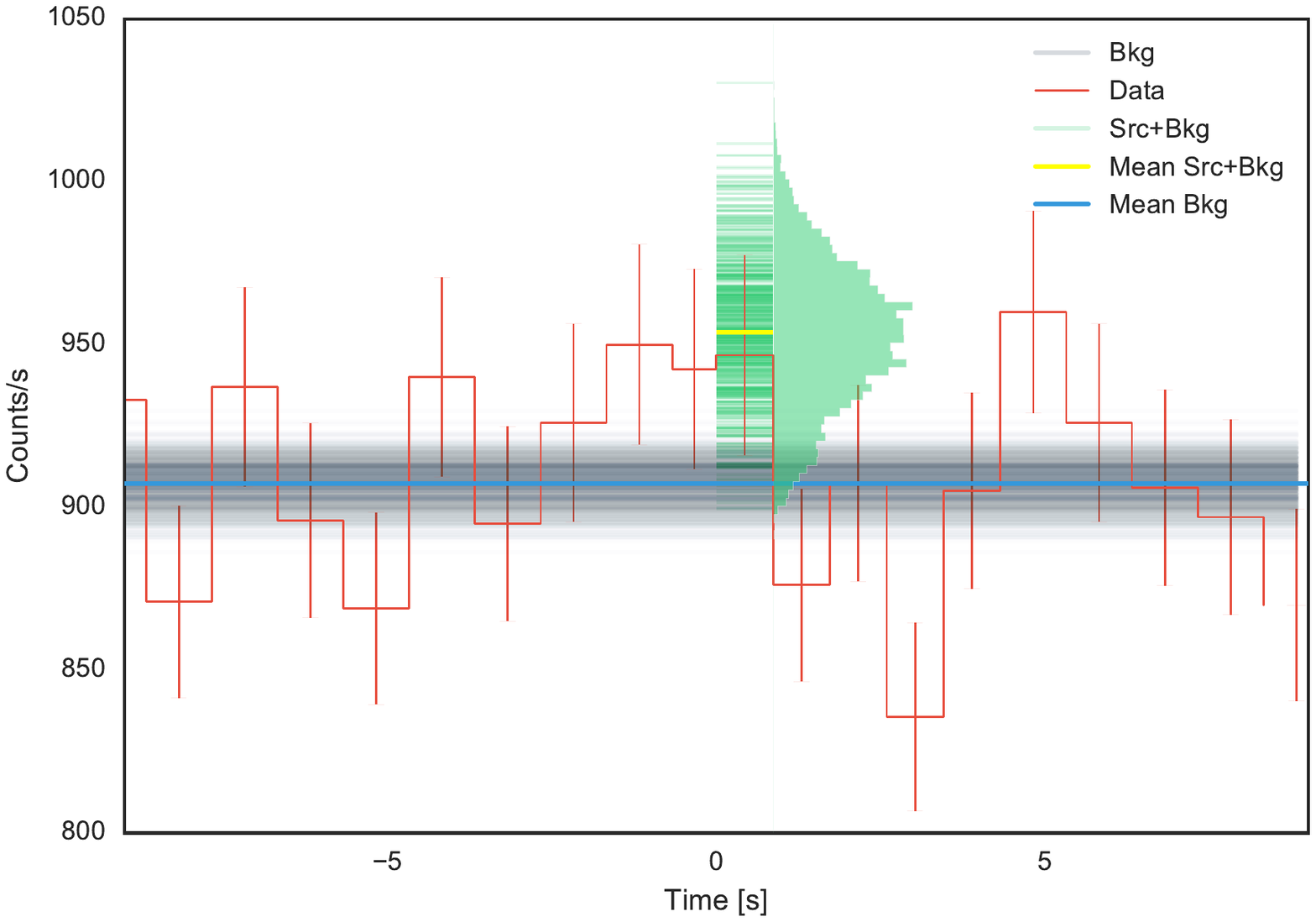}
\caption{The count rate from GRB 140516A in NaI 01 super-imposed with the
source+background and background posteriors. The DIC between the two models
favors source + background, despite the fact that in the broad-band lightcurve
the GRB is hardly evident.}
\label{fig:GRB120212}
\end{center}
\end{figure}

\begin{figure}[htbp]
\begin{center}
\vspace{-0.3cm}
\includegraphics[width=11cm]{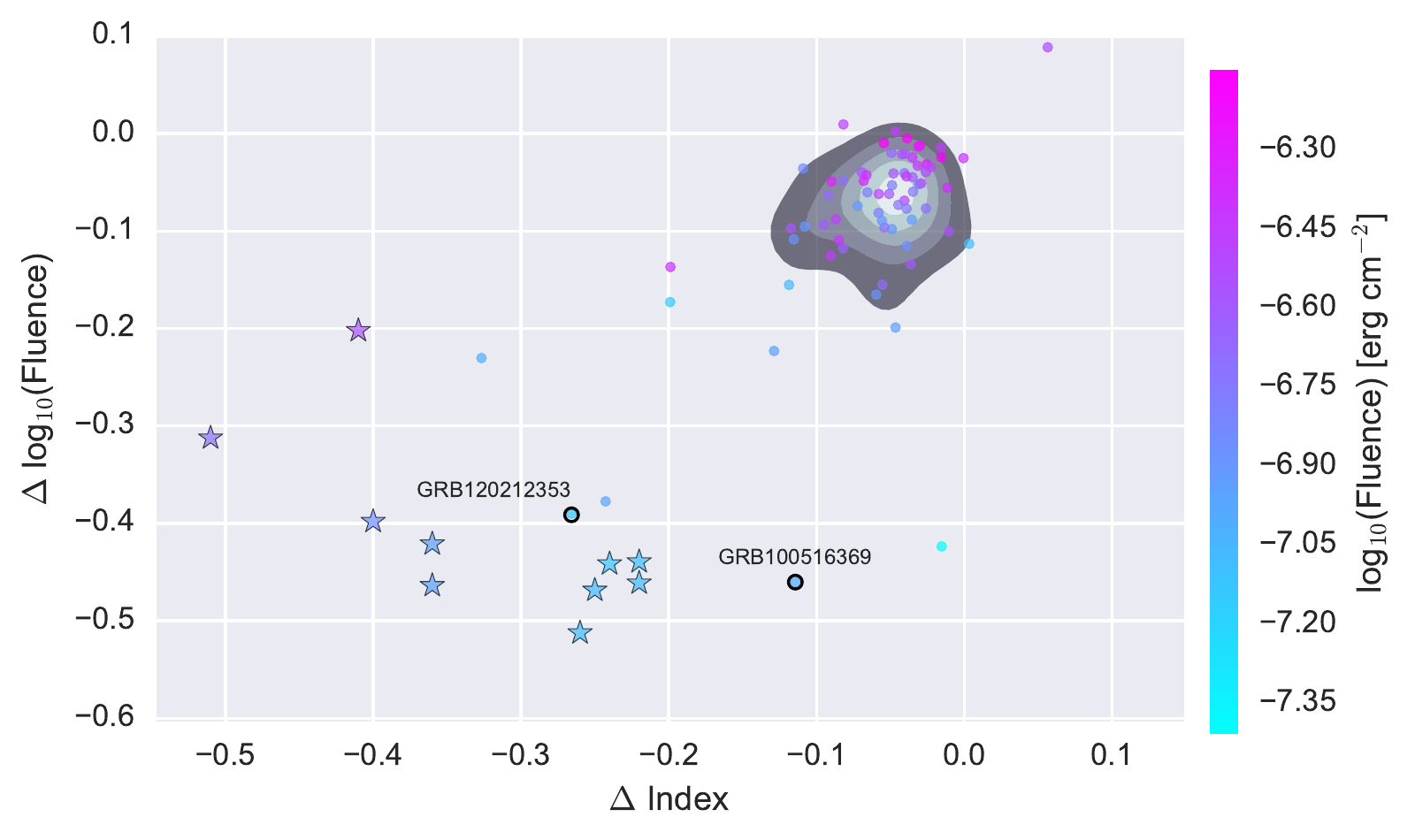}
\caption{The differences in spectral slope and fluence between
  analyses using either RMFIT or PGStat, color-coded according to the
  significance of the residual source counts, and using all short-duration
  GRBs from \cite{Gruber:2012}. For bright (short) GRBs the differences
  in the two approaches are minimal, and within the 1$\sigma$ errors of
  the parameter estimates. However, the fainter the bursts, the larger 
  are the changes towards fainter flux and softer spectral slope.
  The shaded contours contain the $\approx$80\% brightest short GRBs
  (each dot represents one GRB). For two of the faintest short GRBs
  (encircled and labeled dots),
  a preliminary consistency check of the
  PGStat-based parameters with the expected significance in INTEGRAL/ACS
  finds perfect matches. 
  A more thorough cross-check is presently being performed as a combined
  effort of the Fermi/GBM and INTEGRAL/ACS teams which will be published 
  separately.
  The filled stars represent our PGStat-based
  parameters for the GBM event at the 10 locations along the LIGO arc.
  }
\label{fig:deltas}
\end{center}
\end{figure}

We can take it yet another step further and test
the validity of the PGStat analysis using GBM
sub-threshold short GRBs, i.e. GRBs which were detected by Swift/BAT, 
but not by GBM. We pick two GRBs (one faint and one stronger) 
out of the 7 for which \cite{Burns:2016}
determined visibility by GBM, and which
were recovered in ground-based searches of the untriggered GBM data 
(see Table 4 of \citealt{Burns:2016}). For these bursts, the
PGStat analysis results in slightly softer and dimmer spectra than the analysis
with RMFIT, consistent with the above analysis. 
The uncertainties in the fluence are small, and  exclude zero fluence,
i.e. the result recovers proper flux from these two GRBs. This shows that the 
PGStat analysis is capable of recovering real GBM sub-threshold events. 
In contrast,  PGStat analysis of GW150914-GBM indicates an amplitude 
consistent with zero.

To further validate the Bayesian method, we apply it to one of these
weaker GRBs 
(GRB 140516A). The method is able to clearly distinguish between source 
and background (Fig. \ref{fig:GRB120212})
with a DIC preferring a source even though the total count rate 
above background is relatively low (even lower then the GBM-GW150914!). 
This further strengthens our claim that were GBM-GW150914 a real astrophysical
short GRB, the method would be able to identify it. Caution must be taken 
when assessing the presence of a source based on total count rate alone when 
spectral information is available. The additional spectral information, 
as well as the variability of the background, place severe constraints 
on the presence of weak events.

\section{INTEGRAL detectability}
\label{sub:INTEGRAL}

Using the best-fit power law models as derived within RMFIT 
(Tab. \ref{tab:fit}), we compute the fluence over the detection 
range of 50 keV to 4.7 MeV (as Fig. 5 in \citealt{2016arXiv160203920C}
and consistent with their sect. 3.2).

We then use these position-dependent spectral parameters
to simulate the response of ACS for each of these 10 positions,
and find that the GBM event should have been detected at $>$8$\sigma$
everywhere (see Tab. \ref{ACS}). The last column gives the range 
of significances when accounting for both, the error in spectral
slope as well as in amplitude: the smallest significance is still
larger than$>$5.9$\sigma$.

\begin{figure}[ht]
\centering
\includegraphics[width=10.2cm]{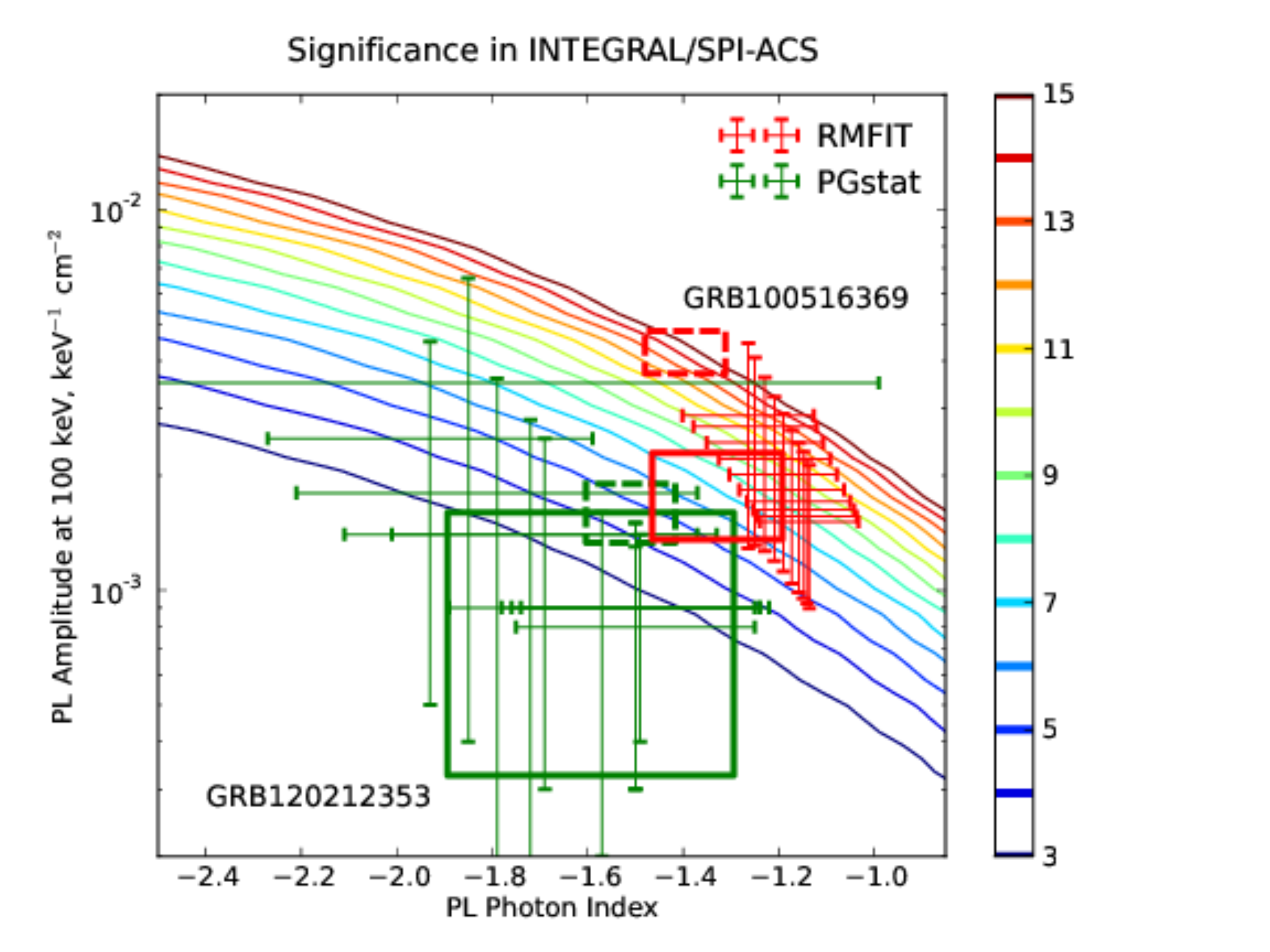}
\caption{
INTEGRAL/SPI-ACS significance map for power law
  spectra of different slopes from the best-fit position on the LIGO arc
  as given in \cite{2016arXiv160203920C}. With the fluences for a 1 sec
  transient as given in Tabs. 1 and 2, the green and red crosses show
  the predicted significances of detection for the claimed GBM event.
  The rectangles are the faint short GRBs used for 
  verification of the PGStat parameters, and are detected with INTEGRAL/ACS
  (we note that the significance contours are only approximate for the two 
  real GRBs, and different for each, due to different incidence angle/response
  and duration).
  \label{fig:ACSmap}}
\end{figure}

\begin{table}[hb]
  \caption{ACS response to GBM event along the LIGO arc, based on the spectra
  as deduced with RMFIT, shown in Tab. \ref{tab:fit}.}
  \centering
  \begin{tabular}{rccc}
    \hline
    \noalign{\smallskip}
   RA / Decl.  & Fluence        & Significance & Significance\\
     (2000.0)  & (50--4700 keV) & of best-fit  & range \\
               & (erg/cm$^2$)     & (sigma)        & (sigma) \\
     \noalign{\smallskip}
     \hline
     \noalign{\smallskip}
  84.0~~  --72.8 & 5.71$\times10^{-7}$ & 12.1 & 8.1 -18.1\\
 155.3~~  --43.2 & 5.39$\times10^{-7}$ & ~8.0 & 5.9 -10.9\\
 102.0~~  --73.9 & 4.96$\times10^{-7}$ & 11.7 & 7.9 -17.0\\
 118.3~~  --72.9 & 4.58$\times10^{-7}$ & 10.9 & 7.6 -15.7\\
 132.0~~  --70.4 & 4.26$\times10^{-7}$ & 10.3 & 7.2 -14.5\\
 140.9~~  --66.6 & 3.98$\times10^{-7}$ & ~9.7 & 6.9 -13.5\\
 147.5~~  --62.5 & 3.79$\times10^{-7}$ & ~9.1 & 6.6 -12.6\\
 151.2~~  --58.0 & 3.66$\times10^{-7}$ & ~8.7 & 6.3 -12.0\\
 153.4~~  --53.1 & 3.55$\times10^{-7}$ & ~8.5 & 6.1 -11.6\\
 153.9~~  --48.2 & 3.46$\times10^{-7}$ & ~8.2 & 6.0 -11.2\\
  \noalign{\smallskip}
     \hline
  \end{tabular}
  \label{ACS}
\end{table}

In contrast, if we use the spectral parameters of our own spectral analysis
using PGStat, the lower fluences (Tab. \ref{tab:fit}) result 
in predictions of typically $<$3$\sigma$ significance (Fig. \ref{fig:ACSmap}), 
consistent with the non-detection
of the GBM event with INTEGRAL-SPI/ACS \citep{2016arXiv160204180S}.

\section{Conclusions}
\label{sec:conclude}

The event seen by GBM at 0.4 s after the gravitational wave GW150914
is a very faint, sub-threshold event \citep{2016arXiv160203920C}, 
not found by any of the
standard undirected algorithms designed to pick up untriggered events.
Both, the signal-to-noise ratio as well
as the false alarm rate reported in \cite{2016arXiv160203920C} 
depend on the assumed hard spectral shape -- a softer spectral shape
increases the probability of a background fluctuation.

Based on this and our analysis
we conclude that the event \citep{2016arXiv160203920C} seen in GBM 0.4
sec after the gravitational wave event GW\,150914
\citep{2016PhRvL.116f1102A} is very likely not an astrophysical
event due to three reasons: 
(i)   the spectrum is soft, and softer than typical short-hard GRBs,
(ii) the spectrum is consistent with the background spectrum 
 as measured by GBM,
(iii) due to our more thorough background treatment, the fluence 
  of the event is substantially fainter, consistent with zero.
  Combined with the softer slope our spectral deconvolution predicts
  that INTEGRAL-SPI/ACS should not detect this event, consistent with
  reality \citep{2016arXiv160204180S}.

Our conclusion is more consistent with
the expectation of no $\gamma$-ray emission from binary black hole
mergers \citep{Lyutikov:2016}, and requires no effort in
significant fine-tuning of models to explain the alleged delay of the
gamma-ray emission relative to the gravitational wave emission.

We have further identified issues with the standard GBM
fitting tool, RMFIT, in the low-count regime and advocate for careful
study of the statistics when handling the spectra of marginal detections. 
We emphasize that RMFIT can be safely used for events in the 
high-count regime, as has been done in many previous Fermi/GBM publications.
In the low-count regime, the PGStat statistics or our Bayesian method 
properly model the statistical properties of GBM data, and our simulations 
show that in this case, our results are 
more accurate than the statistical tools in RMFIT. 
This has been verified by 
(i) comparing predictions for INTEGRAL/ACS
detectability against its actual performance
for short GRBs, as well as 
(ii) by recognizing faint short GRBs detected by GBM (and confirmed by
INTEGRAL/ACS detections) with even smaller significance in the
 broad-band GBM light curve as the GBM-GW150914 event.
Thus, if GBM-GW150914 were a real astrophysical short GRB, our methods 
would have identified it.

\acknowledgements

The authors acknowledge V. Connaughton for implementing modifications
upon our criticism. 
We are grateful to B. Anderson and J. Conrad (Stockholm University) for sharing 
their statistics expertise with us. JG also thanks K. Arnaud for details on
the Castor statistics description in the XSPEC manual.
This research made use of PyMC3 a probabilistic programming language 
for Bayesian inference (Salvatier et al. 2016), Astropy, a community developed
core Python package for Astronomy (Astropy Collaboration et al. 2013), 
Matplotlib, an open source Python graphics environment (Hunter 2007) and 
Seaborn (Waskom et al. 2015) for plotting.

The Fermi/GBM project is supported by NASA. Support for the German contribution
to GBM was provided by the Bundesministerium f\"ur Bildung und Forschung 
(BMBF) via the Deutsches Zentrum f\"ur Luft- und Raumfahrt (DLR) under
contract number 50 QV 0301. This research was supported by the DFG cluster 
of excellence 'Origin and Structure of the Universe' (www.universe-cluster.de).


\end{document}